\documentclass{PoS}

\usepackage{slashed}

\newcommand{\be}{\begin{equation}}
\newcommand{\ee}{\end{equation}}
\newcommand{\bea}{\begin{eqnarray}}
\newcommand{\eea}{\end{eqnarray}}

\title{Operator Structure of TMDs}

\ShortTitle{Operator Structure of TMDs}

\author{\speaker{P.J. MULDERS}\\
        Nikhef Theory Group and Department of Physics and Astronomy, VU University Amsterdam\\
        De Boelelaan 1081, NL-1081 HV Amsterdam, the Netherlands\\
        E-mail: \email{mulders@few.vu.nl}}
        
\abstract{The focus of this talk is on the transverse components of parton momenta. Like for collinear parton distribution functions (PDFs), we are also in the case of transverse momentum dependent (TMD) PDFs, talking about forward matrix elements. While the collinear PDFs describe only spin-spin correlations, the TMD PDFs (or in short TMDs) include spin-momentum correlations, including also time-reversal-odd (T-odd) correlations. The latter are important in the description of single spin asymmetries. In this way TMDs open up new ways of studying the spin structure of hadrons or they can be used as tools that incorporate hadronic structure also in non-collinear situations. The operator structure of TMDs within QCD, in particular the structure of Wilson lines, is more complex than that for collinear functions leading to various ways of breaking of universality. This breaking of universality, however, can be handled within QCD. 
}

\FullConference{QCD Evolution 2015 -QCDEV2015-\\
		26-30 May 2015\\
		Jefferson Lab (JLAB), Newport News Virginia, USA}

\begin{document}

\section{Introduction}

Parton densities or parton distribution functions (PDFs) and decay functions or parton fragmentation functions (PFFs) are natural ingredients in the factorized expressions of cross sections thought of as an incoherent sum of scattering off the partons, quarks and gluons, in a hadron. At high energies, they incorporate the transiton of hadrons into partons and vice versa. High energy introduces the necessary directionality with lightlike directons, in essence the hadron momenta $P$ and its accompanying lightlike direction $n = P^\prime/P{\cdot}P^\prime$, satisfying $P{\cdot}n = 1$. The lightcone momentum fraction $x$ of a parton, $p = x\,P + \ldots$ is linked to scaling variables, such as $x_B = Q^2/2P{\cdot}q$ in (semi-inclusive) deep inelastic scattering (SIDIS or DIS) or $x = q{\cdot}P^\prime/P{\cdot}P^\prime$ in Drell-Yan scattering. Polarization of the target hadrons described via a density operator parametrized by the spin vector $M\,S = S_L\,P + M\,S_T - M\,S_L\,n$ enables the study of polarized densities. The relevant polarized densities are the unpolarized density denoted $q(x) = f_1^q(x)$, the longitudinal spin density $\Delta q(x) = g_1^q(x)$ and the transverse spin density or transversity $\delta q(x) = h_1^q(x)$. For the final state the transition of quark momentum $k = K/z + \ldots$ to the produced hadron momentum $K$ involves the linking of the fraction $z$ to the appropriate scaling variable, e.g.\ $z = q{\cdot}P/K{\cdot}P$ in SIDIS. Scale dependence and parameters to deal with divergences will also be important~\cite{Collins:2011zzd}, but will not be addressed in this contribution.

Considering partonic transverse momenta, $p = x\,P + p_T + (p^2-p_T^2)\,n$, provides new opportunities and novel densities to be studied, involving transverse momentum dependent (TMD) PDFs depending both on the fraction $x$ and the space-like transverse momentum squared $p_T^2$. Important is the fact that one uses the description at high energies, where the three terms in the parametrization of $p$ turn out to contribute at the end of the calculation at order $Q$, $M$ and $M^2/Q$ respectively. The parton virtualities thus are not relevant in the TMDs, only the transverse momentum squared and depending on the process the directionality of the transverse momentum. Measurements of TMDs don't come for free. They involve azimuthal asymmetries in dedicated final states, involving jet directions or final states that can, for instance, be identified through their specific flavor. Often it requires polarized targets to fix a transverse direction or the use of polarimetry in the final state, e.g.\ using the decay orientation of final states ($\rho$ or $\Lambda$ decays). 

In this contribution, we consider the operator structure of TMDs, starting with expressions for hadronic light-front correlators involving specific matrix elements of quark and gluon fields. Some of the results that I will present have been also presented at the DIS2015 meeting~\cite{Mulders:2015cva} and have been used in the study of universality-breaking in $p_T$-widths~\cite{Boer:2015kxa}.

\section{Hadron correlators} 

The wave functions and spinors or polarizations of partons appear in the single particle matrix elements of the corresponding fields. For a basic definition of parton densities, one needs to consider the matrix elements of the fields between hadronic states, a single hadronic target in inclusive deep inelastic scattering (DIS), which in the cross section leads to forward matrix elements of combinations of field operators. For local operator combinations one can still employ the full field theoretical machinery. Such matrix elements are the moments of parton densities. The kinematics in a high-energy process enables a twist expansion in which the leading local operator combinations are identified, which then through a Mellin transform can be identified with collinear parton distribution functions (collinear PDF's). The renormalization differs for each of the local (composite) operators involving for each of them anomalous dimensions, leading to multiplicative renormalization factors for the moments and a convolution of PDFs and splitting functions to account for the scale dependence of the matrix elements and parton densities.

The quark and gluon TMD correlators in terms of matrix elements of quark fields~\cite{Collins:1981uw,Collins:1981uk} include gauge links or Wilson lines $U$, which are needed for color gauge invariance, particularly relevant in the TMD case. They are given by
\begin{eqnarray}
&&
\Phi_{ij}^{[U]}(x,p_T;n)
=\int \frac{d\,\xi{\cdot}P\,d^{2}\xi_T}{(2\pi)^{3}}
\,e^{ip\cdot \xi} \langle P{,}S\vert\overline{\psi}_{j}(0)
\,U_{[0,\xi]}\psi_{i}(\xi)\vert P{,}S\rangle\,\big|_{LF},
\label{e:operator}
\\&&
2x\,\Gamma^{[U,U^\prime]\,\mu\nu}(x,p_T;n) ={\int}\frac{d\,\xi{\cdot}P\,d^2\xi_T}{(2\pi)^3}\ e^{ip\cdot\xi}
\,\langle P{,}S\vert\,F^{n\mu}(0)\,U_{[0,\xi]}^{\phantom{\prime}}\,F^{n\nu}(\xi)\,U_{[\xi,0]}^\prime\,\vert P{,}S\rangle\big|_{LF}
\end{eqnarray}
(color summation for quarks and color tracing for matrix-valued gluon fields are implicit), where the Sudakov decomposition for the momentum $p^{\mu}$ of the produced quark or gluon is used. The non-locality in the integration is limited to the lightfront, $\xi{\cdot}n = 0$, indicated with LF. The gauge links $U_{[0,\xi]}^{\phantom{\prime}}$ are path ordered exponentials needed to make the correlator gauge invariant~\cite{Belitsky:2002sm,Boer:2003cm}. Depending on the process under consideration different gauge links will appear~\cite{Bomhof:2006dp,Bomhof:2007xt}. For the quark correlator the gauge link bridges the non-locality, which in the case of TMDs involves also transverse separation. The simplest ones are the future- and past-pointing staple links $U_{[0,\xi]}^{[\pm]}$ (or just $[\pm]$) that just connect the points $0$ and $\xi$ via lightcone plus or minus infinity, explicitly $U_{[0,\xi]}^{[\pm]} = U_{[0,\pm\infty]}^{[n]} U_{[0_{\scriptscriptstyle T},\xi_{\scriptscriptstyle T}]}^{{\scriptscriptstyle T}}U_{[\pm \infty,\xi]}^{[n]}$. We use these as the basic building blocks. For matrix-valued gluon fields the most general structure involves two gauge links (in triplet representation), denoted as $[U,U^\prime]$, connecting the positions $0$ and $\xi$ in different ways. The simplest combinations allowed for $[U,U^\prime]$ are $[+,+^\dagger]$, $[-,-^\dagger]$, $[+,-^\dagger]$ and $[-,+^\dagger]$. More complicated possibilities with additional (traced) Wilson loops of the form $U^{[\square]}=U_{[0,\xi]}^{[+]}U_{[\xi,0]}^{[-]}$ = $U_{[0,\xi]}^{[+]}U_{[0,\xi]}^{[-]\dagger}$ or its conjugate are allowed as well. A list with all type of contributions can be found in Ref.~\cite{Buffing:2012sz,Buffing:2013kca}. If $U = U^\prime$ the gauge link corresponds to a single gauge link in the octet representation.

\begin{figure}
\begin{minipage}{0.495\textwidth}
\includegraphics[width=0.98\textwidth]{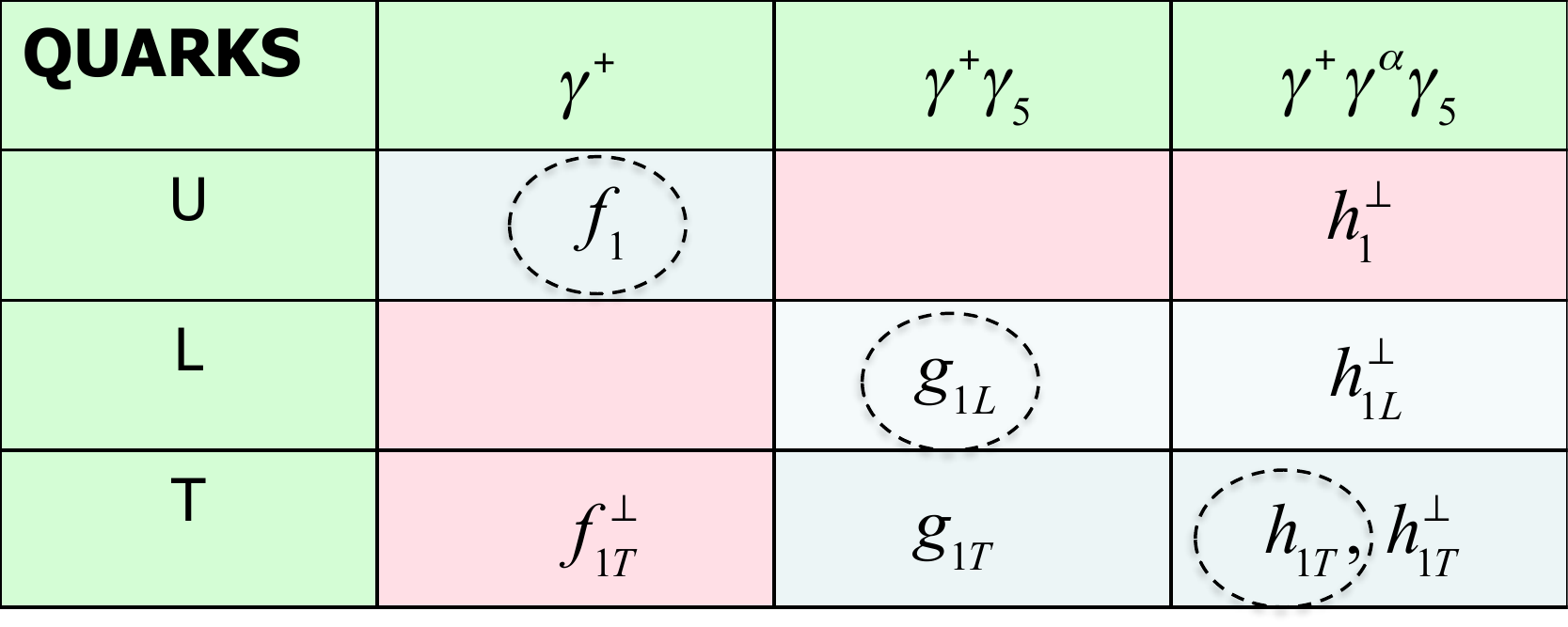}
\end{minipage}
\begin{minipage}{0.495\textwidth}
\includegraphics[width=0.98\textwidth]{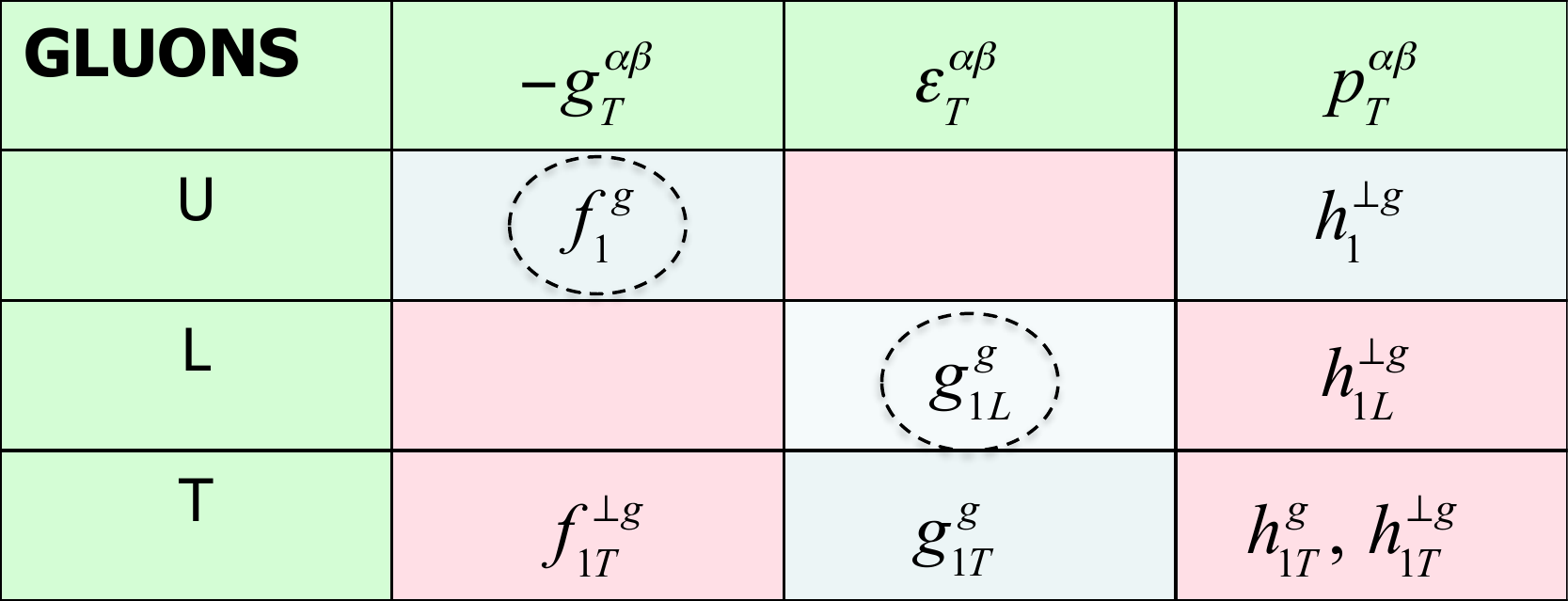}
\end{minipage}
\caption{The Dirac or Lorentz structure in the correlators defines the polarization of the quarks or gluons. The table indicates the TMDs for an unpolarized target ($S_L = S_T = 0$) or for longitudinally and transversed polarized nucleons ($S_L = 1$ or $\vert S_T\vert = 1$). The functions in the pink boxes are T-odd, the circled entries also appear as collinear PDFs (surviving the $p_T$ integration).
}
\label{fig-quarksgluons}
\end{figure}

\begin{figure}
\includegraphics[width=0.48\textwidth]{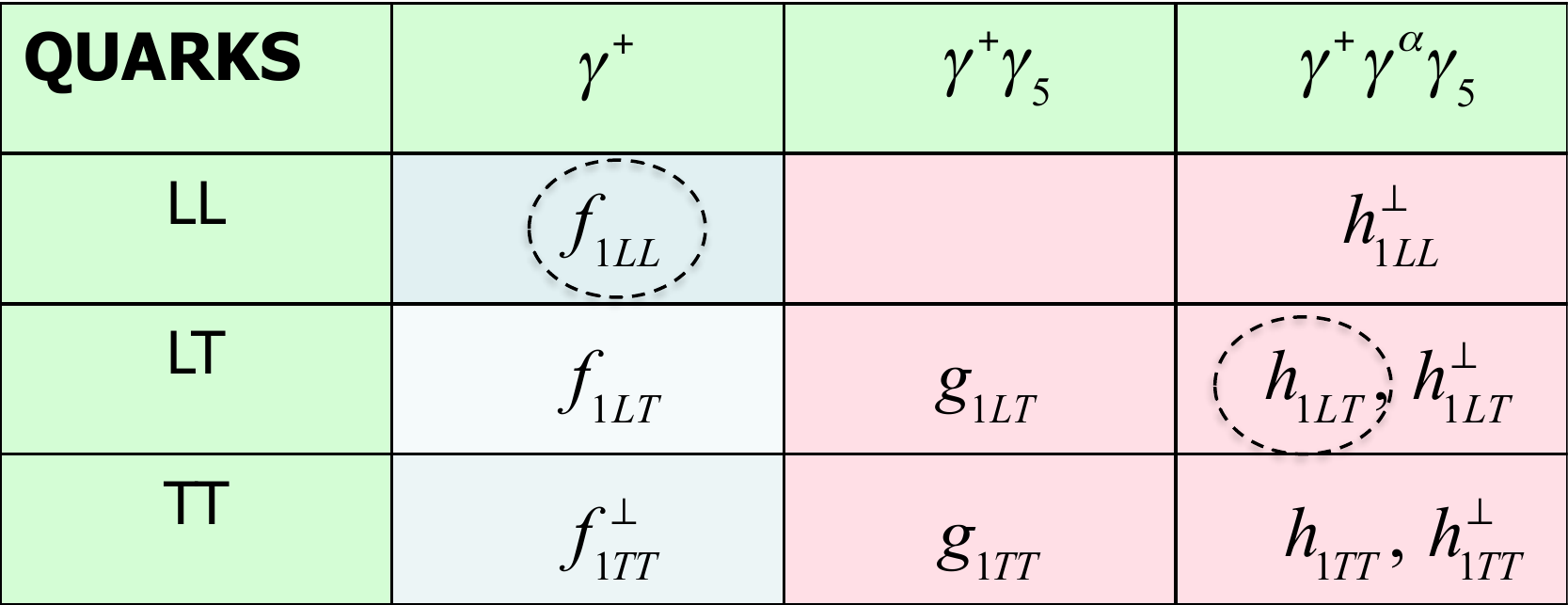}
\caption{The TMDs for a tensor polarized spin one target~\cite{Bacchetta:2000jk}. Again the functions in the pink boxes are T-odd functions. The function $f_{1LL}$ is also known as $b_1$, contributing to the similarly named structure function in scattering off a spin one target~\cite{Hoodbhoy:1988am}. The function $h_{1LT}$ being T-odd was first discussed as distribution function in Ref.~\cite{Bacchetta:2000jk}, while its fragmentation analogue was discussed earlier as a (naturally T-odd) fragmentation function in Ref.~\cite{Ji:1993vw} being named $\hat h_{\bar 1}$ rather than the notation $H_{1LT}$ that we would presently use.
}
\label{quarks-1}
\end{figure}

The above correlators cannot be calculated from first principles and an expansion in terms of TMD PDFs is used, which at the level of leading twist contributions is given by~\cite{Mulders:1995dh,Bacchetta:2006tn,Mulders:2000sh,Meissner:2007rx}
\begin{eqnarray}
\Phi^{[U]}(x,p_{T};n)&=&
f^{[U]}_{1}(x,p_T^2)\frac{\slashed{P}}{2} - f^{\perp [U]}_{1T}(x,p_T^2)\frac{\epsilon_T^{p_T S_T}}{M}\frac{\slashed{P}}{2} + g^{[U]}_{1s}(x,p_T^2)\frac{\slashed{P}\gamma_5}{2} + i\,h_1^{\perp [U]}(x,p_T^2)\frac{[\slashed{p}_T,\slashed{P}]}{4M}
\nonumber \\&&
\mbox{}+h^{[U]}_{1T}(x,p_T^2)\frac{\gamma_5\,[\slashed{S}_{T},\slashed{P}]}{4}
+h_{1s}^{\perp [U]}(x,p_T^2)\,\frac{\gamma_5[\slashed{p}_T,\slashed{P}]}{4M}.
\label{e:QuarkCorr}
\\
2x\,\Gamma^{\mu\nu [U]}(x{,}p_{\scriptscriptstyle T}) &=& 
-g_T^{\mu\nu}\,f_1^{g [U]}(x{,}p_{\scriptscriptstyle T}^2)
+g_T^{\mu\nu}\frac{\epsilon_T^{p_TS_T}}{M}\,f_{1T}^{\perp g[U]}(x{,}p_{\scriptscriptstyle T}^2)
\nonumber\\&&
\mbox{}+i\epsilon_T^{\mu\nu}\;g_{1s}^{g [U]}(x{,}p_{\scriptscriptstyle T})
+\bigg(\frac{p_T^\mu p_T^\nu}{M^2}\,{-}\,g_T^{\mu\nu}\frac{p_{\scriptscriptstyle T}^2}{2M^2}\bigg)\;h_1^{\perp g [U]}(x{,}p_{\scriptscriptstyle T}^2)
\nonumber\\ &&
\mbox{}-\frac{\epsilon_T^{p_T\{\mu}p_T^{\nu\}}}{2M^2}\;h_{1s}^{\perp g [U]}(x{,}p_{\scriptscriptstyle T})
-\frac{\epsilon_T^{p_T\{\mu}S_T^{\nu\}}{+}\epsilon_T^{S_T\{\mu}p_T^{\nu\}}}{4M}\;h_{1T}^{g[U]}(x{,}p_{\scriptscriptstyle T}^2).
\label{e:GluonCorr}
\end{eqnarray}
We have used that $S^\mu = S_{\scriptscriptstyle L}P^\mu + S^\mu_{\scriptscriptstyle T} + M^2\,S_{\scriptscriptstyle L}n^\mu$. For function like $g_{1s}^{[U]}$ and $h_{1s}^{\perp[U]}$ the shorthand notation
\begin{equation}
g_{1s}^{[U]}(x,p_{\scriptscriptstyle T})=S_{\scriptscriptstyle L} g_{1L}^{[U]}(x,p_{\scriptscriptstyle T}^2)-\frac{p_{\scriptscriptstyle T}\cdot S_{\scriptscriptstyle T}}{M}g_{1T}^{[U]}(x,p_{\scriptscriptstyle T}^2)
\end{equation}
is used.
For later use we note that one can use the irreducible (traceless) tensors such as for rank two the tensor $p_T^{\alpha\beta} \equiv p_T^\alpha p_T^\beta+\frac{1}{2}g_T^{\alpha\beta}p_T^2$ and rewrite the corresponding part
\bea
&&
h^{[U]}_{1T}(x,p_T^2)\frac{\gamma_5\,[\slashed{S}_{T},\slashed{P}]}{4}
+h_{1s}^{\perp [U]}(x,p_T^2)\,\frac{\gamma_5[\slashed{p}_T,\slashed{P}]}{4M} \nonumber 
\\ &&\quad \mbox{} =
h^{[U]}_{1}(x,p_T^2)\frac{\gamma_5\,[\slashed{S}_{T},\slashed{P}]}{4}
+S_L h_{1L}^{\perp [U]}(x,p_T^2)\,\frac{\gamma_5[\slashed{p}_T,\slashed{P}]}{4M}
-h_{1T}^{\perp [U]}(x,p_T^2)\,\frac{p_{T\alpha\beta}S_T^{\{\alpha}\gamma_T^{\beta\}}\slashed{P}\gamma_5}{4M^2} 
\eea
with $h^{[U]}_1 = h^{[U]}_{1T} - (p_T^2/2M^2)\,h_{1T}^{\perp [U]}$ absorbing a trace part.
The gauge link dependence in this parametrization is at this point still contained in the TMDs and indicated with the superscripts $[U]$. Note that for quarks $f_{1T}^{\perp}$ and $h_1^\perp$ are T-odd, while for gluons $f_{1T}^{\perp g}$, $h_{1T}^{g}$, $h_{1L}^{\perp g}$ and $h_{1T}^{\perp g}$ are T-odd. The TMDs and their polarization properties are summarized in Fig.~\ref{fig-quarksgluons} for quarks and gluons in a spin 1/2 target. In a spin 1 target there would be besides the functions that also would appear in a spin 1/2 target (unpolarized or vector polarized) in addition other TMD functions for a tensor polarized target~\cite{Bacchetta:2000jk}, illustrated in Fig.~\ref{quarks-1}.


\section{Gauge links and color flow}

Even if any gauge link defines a gauge invariant correlator, the relevant gauge links to be used for a particular correlator used in a given process just results from a correct resummation of all diagrams including the exchange of any number of $A^n$ (or $A^+$) gluons between the hadronic parts and the hard part, i.e.\ gluons with their polarization along the hadronic momentum. They nicely sum to the path-ordered exponential. For quark distributions in semi-inclusive deep inelastic scattering they resum into a future-pointing gauge link, in the Drell-Yan process they resum into a past-pointing gauge link, which is directly linked to the color flow in these processes. This color flow dependence has extensively been discussed in Refs~\cite{Bomhof:2006dp,Bomhof:2007xt}.

\begin{figure}[t]
\begin{minipage}{0.51\textwidth}
\includegraphics[width=0.95\textwidth]{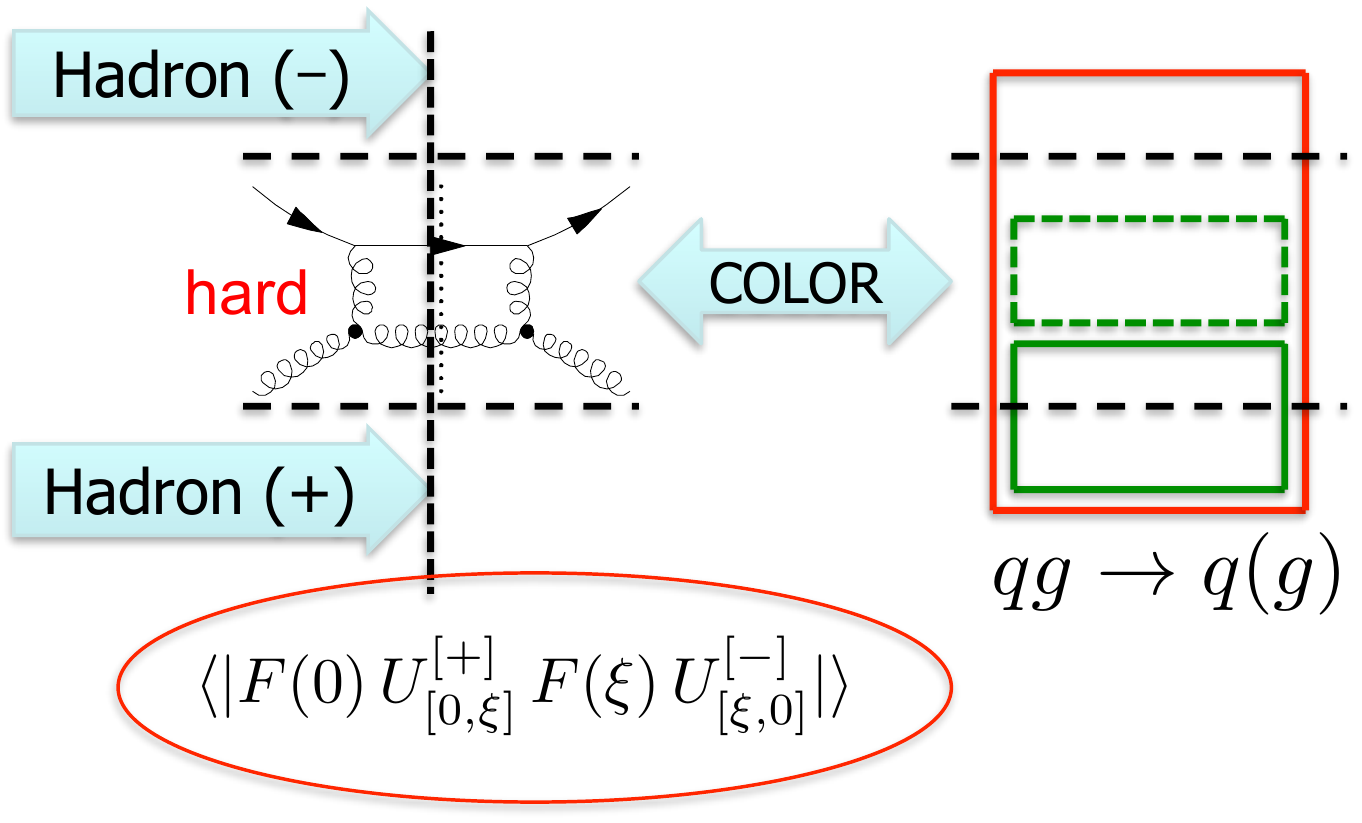}
\end{minipage}
\begin{minipage}{0.48\textwidth}
\caption{This figure illustrates the case of a gluon correlator for hadron (+). One of the possible color flows is indicated in the right half of the figure. The (green) color-line running from hadron (+) into the final state leads to the future pointing link $U^{[+]}_{[0,\xi]}$. The (red) color-line running via the initial state hadron (-) leads to the past-pointing link $U^{[-]}_{[\xi,0]}$. The (green) dashed line has a $U^{[+]}U^{[+]\dagger}$ structure and disappears. This results into the indicated gauge link structure. For the quark correlator in hadron (-) one has a $U^{[+]}$ gauge link structure. 
}
\label{colortrack}
\end{minipage}
\end{figure}

\begin{figure}
\begin{center}
\includegraphics[width=0.7\textwidth]{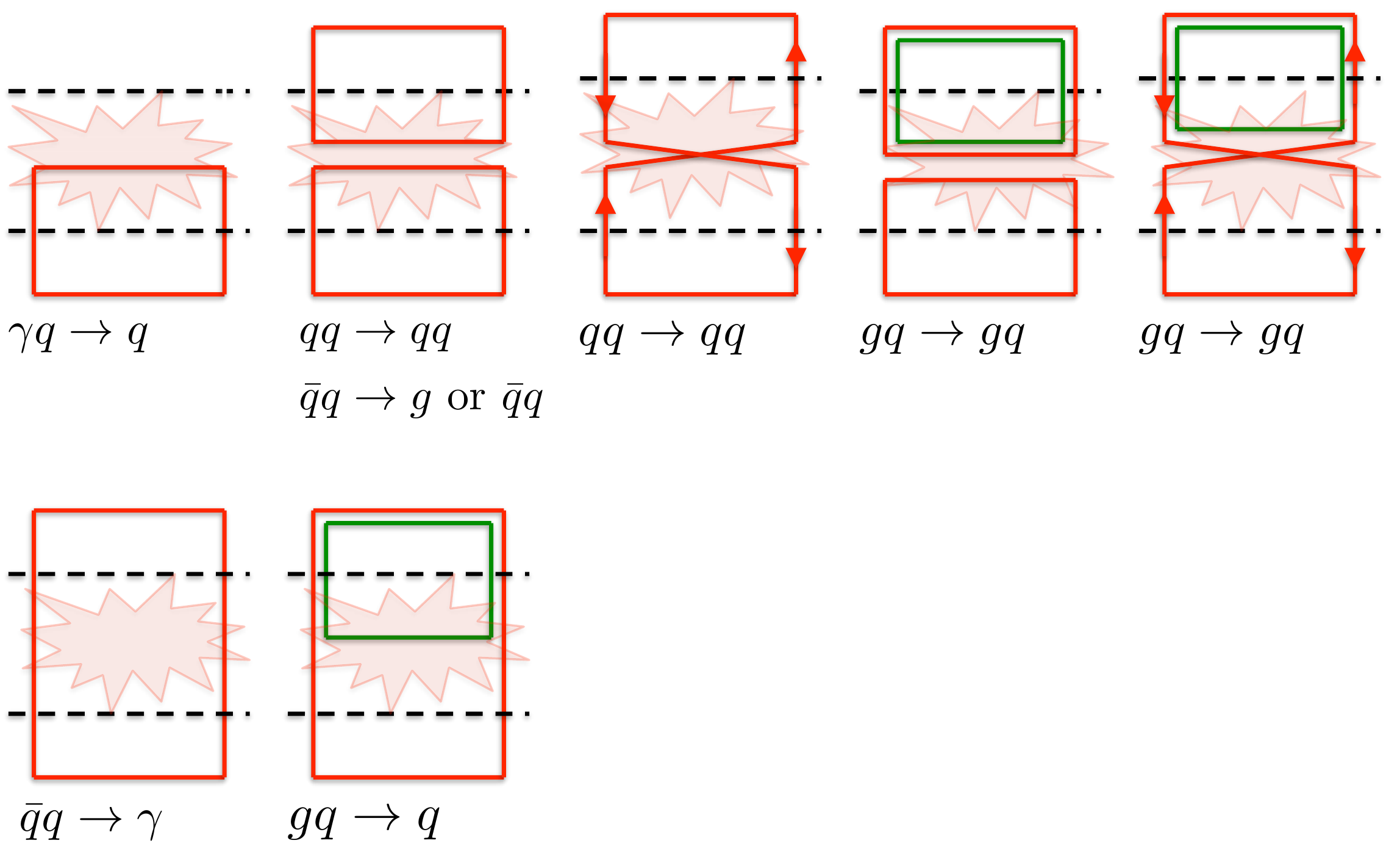}
\end{center}
\caption{Color flow classes relevant for a quark correlator in the lower hadron and examples of processes where they appear. The flows on the left lead to the structures $\langle \overline\psi(0)U^{[+]}_{[0,\xi]}\psi(\xi)\rangle$ (upper left) and $\langle \overline\psi(0)U^{[-]}_{[0,\xi]}\psi(\xi)\rangle$ (lower left). The second flow in the upper row has an additional traced Wilson loop, i.e.\ $\langle \overline\psi(0)U^{[+]}_{[0,\xi]}{\rm Tr}(U^{[-]}_{[\xi,0]}U^{[+]}_{[0,\xi]})\psi(\xi)\rangle$, while the third has a structure with additional winding, $\langle \overline\psi(0)U^{[+]}_{[0,\xi]}U^{[-]}_{[\xi,0]}U^{[+]}_{[0,\xi]}\psi(\xi)\rangle$, etc.
}
\label{flowquarks}
\end{figure}

\begin{figure}
\begin{center}
\includegraphics[width=0.7\textwidth]{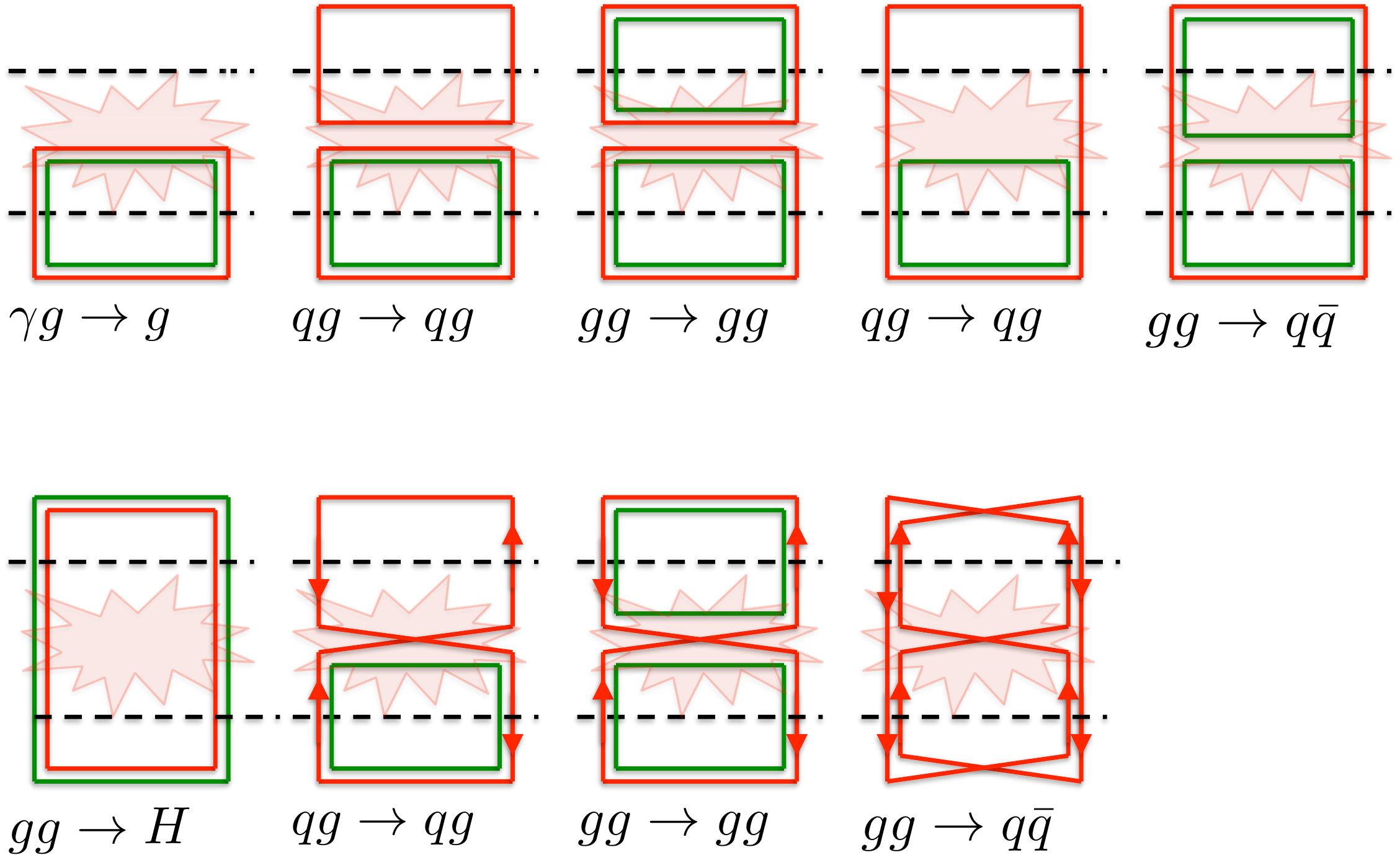}
\end{center}
\caption{Color flow classes relevant for a gluon correlator in the lower hadron and examples of processes where they appear. The flows on the left lead to the simplest gauge link structures $\langle F(0)U^{[+]}_{[0,\xi]}F(\xi)U^{[+]}_{[\xi,0]}\rangle$ (upper left) and $\langle F(0)U^{[-]}_{[0,\xi]}F(\xi)U^{[-]}_{[\xi,0]}\rangle$ (lower left). Note that for a given diagram, such as the $qg\rightarrow qg$ contribution, 
multiple color flow possibilities exist.
}
\label{flowgluons}
\end{figure}

We illustrate the main feature in Fig.~\ref{colortrack} for the gauge link structure in a gluon correlator. It looks like this will lead to an enormous complexity, but the fact that color structure into the final state does not give rise to universality issues and that loops fully contained in the hard part (like the dashed loop in Fig.~\ref{colortrack}) are irrelevant, limits the possibilities to just a limited set of possibilities, illustrated for quark and gluon correlators in Figs~\ref{flowquarks} and \ref{flowgluons}. We do note that in a given hard process, many color flow possibilities will enter as is for instance known from diagrammatic methods to determine color factors. For instance in the $qg \rightarrow qg$ amplitude three different classes enter. Each diagram will have color flow possibilities leading to factors that combine into the color factor for a particular process, like the color factor 1 in DIS, the color factor $1/N_c$ in Drell-Yan and the color factor $N_c$ for quark-antiquark production processes. In this case one may collect for a correlator with a particular gauge link a different combination from the various color flow possibilities, which as expected also turn out to be gauge invariant as noted in Ref.~\cite{Bomhof:2006ra}. 

\section{Operator analysis}

In the situation of collinear PDFs (integrated over transverse momenta), the non-locality is restricted to the lightcone, $\xi{\cdot}n = \xi_T = 0$ (LC) and the staple links reduce to straight-line Wilson lines. The correlators then involve the non-local operator combinations $\overline\psi(0)U^{[n]}_{[0,\xi]}\psi(\xi)\vert_{LC}$ or $F^{n\mu}(0)U^{[n]}_{[0,\xi]}F^{n\nu}(\xi)U^{[n]}_{[\xi,0]}\vert_{LC}$, expanded in terms of local leading twist operators $\overline\psi(0)D^n\ldots D^n\psi(0)$ and ${\rm Tr}[F^{n\mu}D^n\ldots D^n F^{n\nu}(0)D^n\ldots D^n]$. The local operators are found by taking $x$-moments of the collinear functions. These collinear functions are indicated in Figs~\ref{fig-quarksgluons} and \ref{quarks-1} as entries circled by dashed lines. Although gauge links are part of the matrix elements, they do in that case not cause any non-universality or process dependence. In order to find the local operators for TMDs we integrate over transverse momentum including explicit transverse momentum vectors as weights. Each transverse momentum $p_T^\alpha$ in the weight becomes a derivative with transverse index. The simplest of these transverse moments are
\begin{eqnarray}
&&
\int d^2p_{\scriptscriptstyle T}\ \Phi^{[U]}(x,p_{\scriptscriptstyle T})=\widetilde\Phi(x),
\label{ptmoment-0}
\\&&
\int d^2p_{\scriptscriptstyle T}\ p_{\scriptscriptstyle T}^{\alpha}\,\Phi^{[U]}(x,p_{\scriptscriptstyle T})=\widetilde\Phi_{\partial}^{\alpha}(x)+C_{G,c}^{[U]}\,\widetilde\Phi_{G,c}^{\alpha}(x),
\label{ptmoment-1}
\\&&
\int d^2p_{\scriptscriptstyle T}\ p_{\scriptscriptstyle T}^{\alpha_1}p_{\scriptscriptstyle T}^{\alpha_2}\,\Phi^{[U]}(x,p_{\scriptscriptstyle T})=\widetilde\Phi_{\partial\partial}^{\alpha_1\alpha_2}(x)+C_{G,c}^{[U]}\,\widetilde\Phi_{\{\partial G\},c}^{\alpha_1\alpha_2}(x)+C_{GG,c}^{[U]}\,\widetilde\Phi_{GG,c}^{\alpha_1\alpha_2}(x),
\label{ptmoment-2}
\end{eqnarray}
and similarly results for $\Gamma^{[U,U^\prime]}(x,p_T)$. These integrated results will require standard UV regularization and corresponding scale dependence, while one also may need to consider appropriate combinations, e.g.\ subtraction of traces, to get finite results. Often it is more appropriate to work with Bessel moments~\cite{Boer:2011xd}. The important point in Eqs~\ref{ptmoment-0} to \ref{ptmoment-2} is that the correlators appearing in these moments are of the form 
\bea
&&
\widetilde\Phi_{\hat O,ij}^{[U]}(x,p_{T})
=\int \frac{d\,\xi{\cdot}P\,d^{2}\xi_{T}}{(2\pi)^{3}}
\,e^{ip\cdot \xi} \langle P{,}S\vert\overline{\psi}_{j}(0)
\,U_{[0,\xi]}\hat O(\xi)\psi_{i}(\xi)\vert P{,}S\rangle\,\Big|_{LF},
\label{e:twistoperatorquark}
\eea
with different types of operators $\hat O(\xi)$ built from $i\partial_T(\xi) = iD_T^\alpha(\xi) - A_T^\alpha(\xi)$ and $G^\alpha(\xi)$. These are defined in a color gauge invariant way (thus including gauge links),
\bea
&&A_{T}^{\alpha}(\xi)=\frac{1}{2}\int_{-\infty}^{\infty}
d\eta{\cdot}P\ \epsilon(\xi{\cdot}P-\eta{\cdot}P)
\,U_{[\xi,\eta]}^{[n]} F^{n\alpha}(\eta)U_{[\eta,\xi]}^{[n]}, 
\label{e:defA} 
\\
&&G^{\alpha}(\xi)=\frac{1}{2}\int_{-\infty}^{\infty}
d\eta{\cdot}P\ U_{[\xi,\eta]}^{[n]}F^{n\alpha}(\eta)
U_{[\eta,\xi]}^{[n]},
\label{e:defG}
\eea
with $\epsilon (\zeta)$ being the sign function. Note that $G^{\alpha}(\xi)$ = $G^{\alpha}(\xi_T)$ does not depend on $\xi{\cdot}P$, implying in momentum space $p\cdot n = p^+ = 0$, hence the name gluonic pole matrix elements~\cite{Efremov:1981sh,Efremov:1984ip,Qiu:1991pp,Qiu:1991wg,Qiu:1998ia,Kanazawa:2000hz}. In the $p_T$ moments one encounters symmetrized products of these operators indicated with subscripts $\{\partial G\}$, etc. Moreover the color summation often introduces multiple possibilities, e.g.\ for a gluonic pole matrix element in combination with two gluon fields there are two possibilities, ${\rm Tr}(F[G,F])$ ($c=1$) and ${\rm Tr}(F\{G,F\})$ ($c = 2$) that have to be summed over. The final important ingredient in the $p_T$ moments are the gluonic pole factors, calculable factors depending on the number of gluonic poles in the operator and the path of the gauge link $U$. Most well-known are the single gluonic pole factors $C_G^{[\pm]} = \pm 1$. Other examples are given in Table~\ref{t:gpfactors} for quarks~\cite{Buffing:2012sz} or in Ref.~\cite{Buffing:2013kca} for gluons.

\begin{table}[t]
\begin{center}\begin{tabular}{|r||c|c|c|}
\hline
$\mbox{}\quad\rule{0pt}{4mm}\Phi^{[U]}$
&$\mbox{}\quad\Phi^{[\pm]}\quad\mbox{}$
&$\mbox{}\quad\Phi^{[+\Box]}\quad\mbox{}$
&$\mbox{}\ \ \Phi^{[(\Box)+]}\ \ \mbox{}$
\\[2pt]
\hline
\hline
$C_{G}^{[U]}$&
$\pm1$&$3$&$1$
\\[1pt]
\hline
$C_{GG,1}^{[U]}$&
$1$&$9$&$1$
\\[1pt]
\hline
$C_{GG,2}^{[U]}$&
$0$&$0$&$4$
\\[1pt]
\hline
\end{tabular}\end{center}
\caption{
The values of the gluonic pole pre-factors for some gauge links needed in the $p_T$-weighted cases for quark correlators. The index $[+\Box]$ indicates a gauge link $U^{[+]}U^{[\Box]}$ while  $[+(\Box)]$ indicates the gauge link $U^{[+]}$\,Tr$(U^{[\Box]})/N_c$. Note that the value of $C_{G}^{[U]}$ is the same for single and double transverse weighting, i.e.\ $C_{\{\partial G\}}^{[U]}$ = $C_{G}^{[U]}$ They just depends on the number of gluonic poles.
\label{t:gpfactors}}
\end{table}

The operators $\widetilde\Phi_{\hat O}(x)$ are linked to three-parton quark-gluon-quark correlators $\Phi_{D}(x,y)$ with an operator structure $\langle\overline\psi(0)D^\alpha(\eta)\psi(\xi)\rangle$, non-local along the lightcone, and similarly $\Phi_{F}(x,y)$ or gluon-gluon-gluon correlators like $\Gamma_{F}(x,y)$. These multi-parton distributions play a role in higher twist contributions to the cross sections and can be used to establish relations, they appear in sum rules, they exhibit symmetries between correlators involving partons and anti-partons, etc. The specific matrix elements needed in TMD physics are zero momentum limits or integrations over one of the momenta in the multi-parton correlators. The multi-parton distributions often also are driving the large $p_T$-behavior of the TMDs. For the collinear PDFs extended to the corresponding TMDs, this is the $\alpha_s/p_T^2$ behavior of $f_1(x,p_T^2)$, involving the splitting functions for the evolution. Multi-parton distributions are driving the large $p_T$ behavior of many of the TMDs~\cite{Bacchetta:2008xw,Echevarria:2015uaa}, in particular the T-odd ones. Note, however, that the large $p_T$ behavor of TMDs without a collinear counterpart may be driven by collinear functions, such as $f_1^g$ driving the linearly polarized gluon distribution $h_1^{\perp g}$~\cite{Catani:2010pd,Catani:2011kr} or the transversity $h_1$ driving the Pretzelocity function $h_{1T}^\perp$~\cite{Bacchetta:2008xw}.  

\section{Universal TMDs}

Taking transverse derivatives gives the coefficients in the expansion in transverse momenta, for which we like to use traceless irreducible tensors $p_T^{\alpha_1\alpha_2\ldots}$ of a fixed rank, which describe in essence the azimuthal dependence. We use the moments to identify the coefficients in the azimuthal expansion
\begin{equation}
\Phi(x,p_T) = \widetilde\Phi(x,p_T^2) + \frac{p_{Ti}}{M}\,\widetilde\Phi^{i}(x,p_T^2) + \frac{p_{Tij}}{M^2}\,\widetilde\Phi^{ij}(x,p_T^2) + \ldots .
\end{equation}
If in the higher moments just operators of the type $\widetilde\Phi_{\partial\ldots\partial}^{\alpha_1\ldots\alpha_n}$ appear, it is easy to find the operator expressions for $f_1(x,p_T^2)$. It corresponds with the rank zero operator $\widetilde\Phi(x,p_T^2)$ including all $\partial{\cdot}\partial$ traces that are subtracted in the higher rank operators and account for the $p_T^2$ dependence. Such is actually the case for fragmentation functions where the gluonic pole matrix elements (after integration over transverse momenta) vanish~\cite{Gamberg:2008yt,Meissner:2008yf,Gamberg:2010uw}. 

\begin{figure}[t]
\begin{center}
\includegraphics[width=0.8\textwidth]{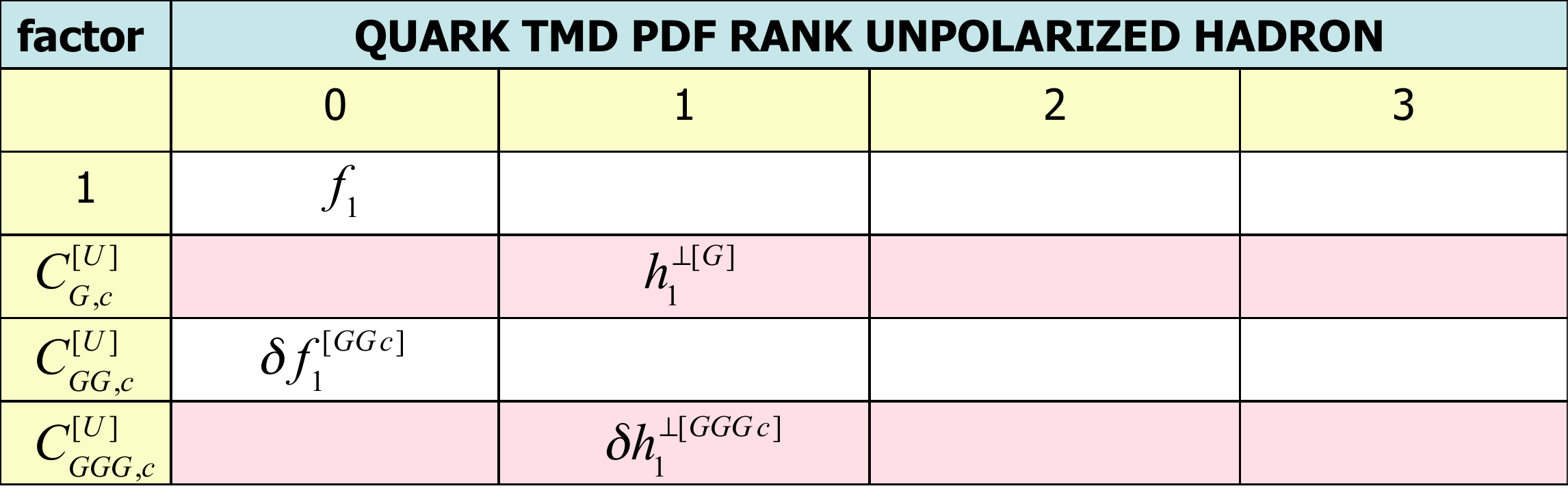}
\\
\includegraphics[width=0.8\textwidth]{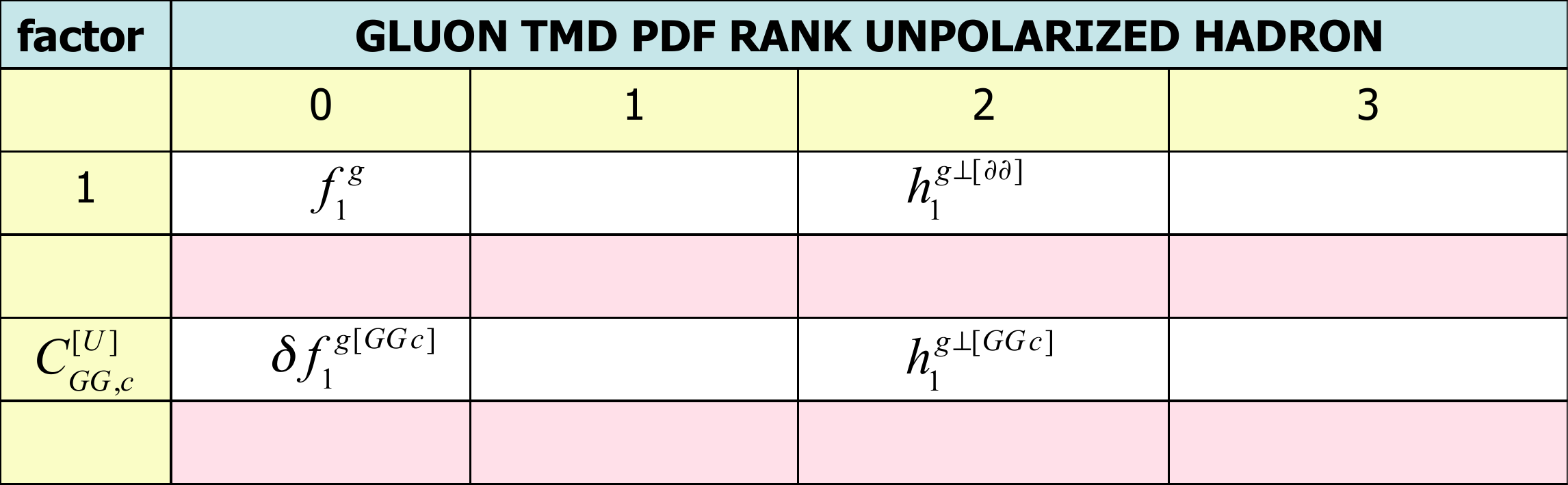}
\end{center}
\caption{Universal set of functions needed for unpolarized quark (upper part of figure) and gluon (lower part of figure) TMDs including up to three gluonic poles. In principle these tables extend to higher number of gluonic poles (vertical range). Horizontally the highest possible rank is limited to $2(S_{\rm parton} + S_{\rm hadron})$, i.e.\ in an unpolarized target to rank one for quark TMDs and rank two for gluon TMDs.
\label{rank-unpolarized}
}
\end{figure}

For the distribution correlators, however, this procedure does not lead to a unique correlator linked to a particular function because gluonic pole matrix elements do not vanish. This leads to two types of $\widetilde\Phi^i$ correlators, $\widetilde\Phi_\partial^I$ and $\widetilde\Phi_G^i$, which easily can be distinguished because they have different time reversal behavior. However, at rank two and higher one also has double gluonic poles of which the trace terms $G{\cdot}G$ lead to new functions. Even the unpolarized quark (or gluon) TMD distributions $f_1(x,p_T^2)$ remain gaugelink-dependent~\cite{Boer:2015kxa} because of this. Only the $\partial{\cdot}\partial$ traces are taken care of in the $p_T^2$-dependence of the function, but gluonic pole trace operators $\widetilde\Phi_{G{\cdot}G,c}(x,p_T^2)$ need to be included and require introduction of functions $\delta f_1^{[GG,c]}(x,p_T^2)$. These functions do not have a collinear equivanlent and must satisfy $\int d^2p_T\ \delta f_1^{[GG,c]}(x,p_T^2) = 0$, so they are responsible for gauge-link dependent modulations in the $p_T^2$ dependence, 
\begin{equation}
f_1^{[U]}(x,p_T^2) = f_1(x,p_T^2) + \sum_{c=1,2}C_{GG,c}^{[U]}\,\delta f_1^{[GG,c]}(x,p_T^2) + \ldots ,
\label{f1full}
\end{equation}
of which only the first term survives in the collinear, $p_T$-integrated, situation. It leads to a process dependent $p_T^2$ behavior, e.g.\ in the $p_T$-width. In the case of the quark correlator are two possible color contractions in the summation over $c$. The functions, their rank and the number of gluonic poles involved in the operator structure for unpolarized hadrons is illustrated in Fig.~\ref{rank-unpolarized}. 

\begin{figure}[t]
\begin{center}
\includegraphics[width=0.8\textwidth]{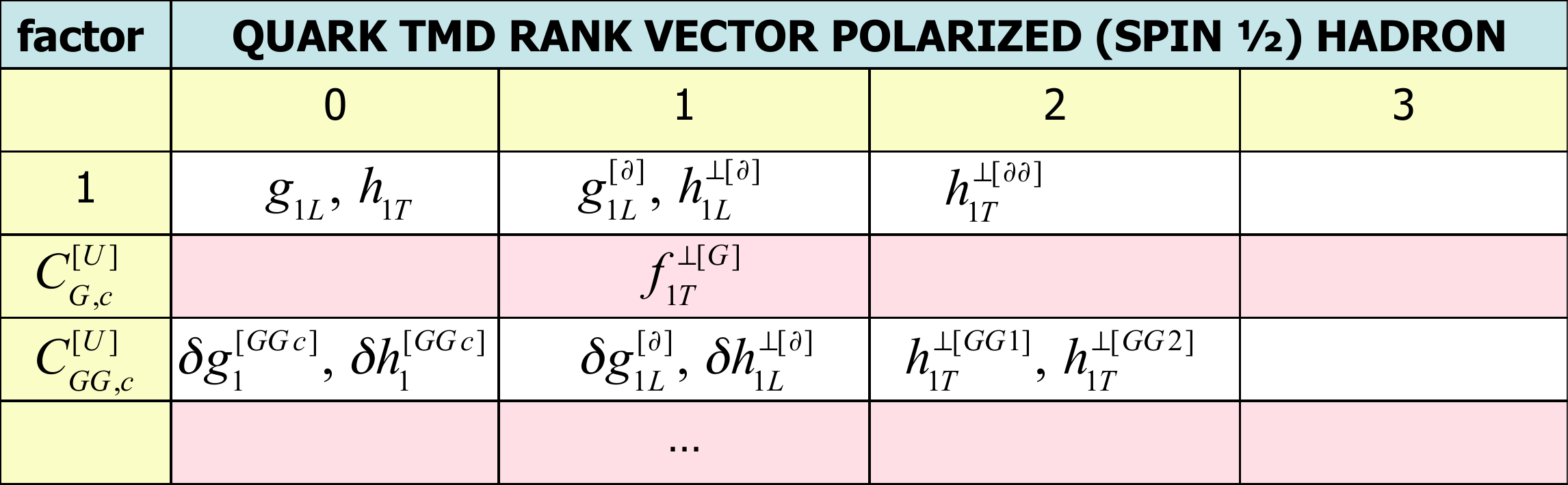}
\\
\includegraphics[width=0.8\textwidth]{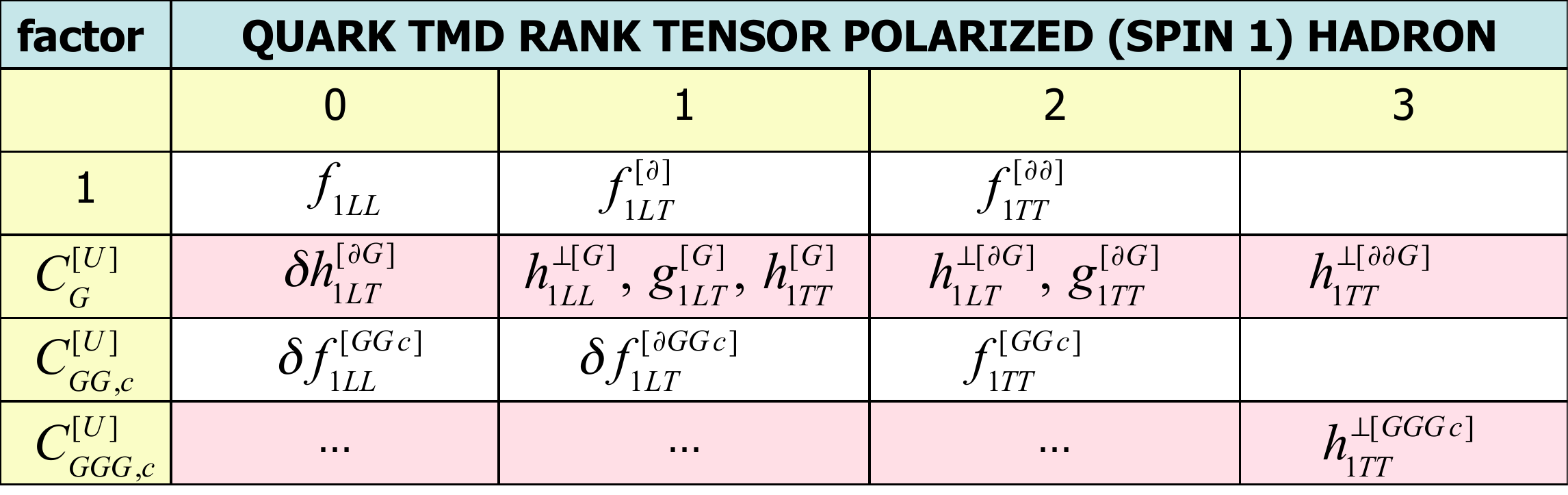}
\end{center}
\caption{Universal set of functions needed for vector and tensor polarized quark TMDs including up to three gluonic poles. For three gluonic poles not all $\delta\Phi$ or $\delta\delta\Phi$ results that can be obtained by adding gluonic pole traces are given, but their construction should be evident.
\label{rank-tensorpolarized}
}
\end{figure}

The results for polarized hadrons are presented in Fig.~\ref{rank-tensorpolarized}. While the first term in Eq.~\ref{f1full} is gaugelink independent, the first term for the Sivers function already has a gluonic pole factor,
\begin{equation}
f_{1T}^{\perp[U]}(x,p_T^2) = C_{G}^{[U]}\,f_{1T}^{\perp[G]}(x,p_T^2) + \ldots .
\end{equation}
For the rank two Pretzelocity distribution there are three different operator structures contributing to the gauge link dependence, 
\begin{equation}
h_{1T}^{\perp[U]}(x,p_T^2) = h_{1T}^{\perp [\partial\partial]}(x,p_T^2) 
+ \sum_{c=1,2}C_{GG,c}^{[U]}\,h_{1T}^{\perp[GG,c]}(x,p_T^2) + \ldots ,
\end{equation}
one of them without gluonic poles and two with gluonic poles. Thus measurements of Pretzelocity effects are gaugelink-dependent, and hence process dependent, even if the observable is a T-even function. In any given process a particular combination of the universal functions on the righthandside appears. These universal functions are here labeled by a combination of $\partial$ and $G$ identifying the operator structure and if needed an index $c$ if multiple color configurations have to be considered. In all of the above expressions, one has to be aware of additional modulations that come from operators with even more (traced) gluonic pole terms. To study their possible importance lattice studies using different gaugelink structures would be useful~\cite{Musch:2011er,Engelhardt:2014wra} as discussed in Ref~\cite{Boer:2015kxa}. 

Turning to the gluon distributions, one also encounters gaugelink-dependence already for the unpolarized TMD distributions,
\begin{eqnarray}
f_{1}^{g[U,U^\prime]}(x,p_T^2)&=&f_1^g(x,p_T^2) 
+ \sum_{c=1}^4 C_{GG,c}^{[U,U^\prime]}\,\delta f_{1}^{g[GG,c]}(x,p_{T}^2) + \ldots, 
\end{eqnarray}
including four different color configurations in the $p_T^2$ modulation. Like for the Pretzelocity, one has for a T-even situation such as that of linearly polarized gluons in an unpolarized hadron also multiple universal functions,
\begin{eqnarray}
h_1^{\perp g[U,U^\prime]}(x,p_T^2)&=&h_1^{\perp g [\partial\partial]}(x,p_T^2)
+\sum_{c=1}^{4}C_{GG,c}^{[U,U^\prime]}\,h_1^{\perp g [U,U^\prime]}(x,p_T^2) + \ldots .
\end{eqnarray}
The above examples illustrate our ongoing efforts~\cite{Buffing:2013kca} to establish a universal set of TMD functions. Without going in more detail here, I want to emphasize that in situations where at least two TMDs with nonzero rank are involved in the initial state (thus hadron-hadron initiated processes), one must account for possible additional color factors in the basic expressions that are in DY-like processes (color neutral final state) for instance different from the $1/N_c$ or $1/(N_c^2-1)$ factors for $q\overline q$ or $gg$ initiated processes~\cite{Buffing:2011mj,Buffing:2013dxa}.

\section{Conclusions}

TMDs provide information on the three-dimensional partonic structure of hadrons. They can in principle be accessed at leading order in hadronic processes provided that one picks the right variable, usually involving azimuthal asymmetries in polarized processes. Such efforts are under investigation in the experimental programs at RHIC/Brookhaven, JLab, BELLE, COMPASS/CERN, JPARC, BESIII or BaBar. Besides experimental efforts,  theoretical developments are underway to understand the data and the way these have to be interpreted in terms of the partonic structure of hadrons. This is a nontrivial enterprise since not only there are many ideas, but also many technical hurdles to take. I have focussed on efforts to establish a universal set of TMDs, connected to specific operators with which one can try to work~\cite{Hautmann:2014kza}. In addition many efforts are ongoing to understand the scale dependence and the matching that is needed to simultaneously understand the behavior at low and high $q_T$ values as well as the links with low-x behavior.

\acknowledgments
This research is part of the research program of the ``Stichting voor Fundamenteel Onderzoek der Materie (FOM)'', which is financially supported by the ``Nederlandse Organisatie voor Wetenschappelijk Onderzoek (NWO)'' and the EU "Ideas" programme QWORK (contract 320389).


\end{document}